\documentclass[12pt]{iopart}
 
\usepackage{graphicx}

\usepackage{amssymb}

\newcommand{\Ai}{\mathrm{Ai}}

\begin{document}

\title{Angular minimum uncertainty states with large uncertainties}

\author{J{\"o}rg B. G{\"o}tte\dag, Paul M. Radmore\ddag, Roberta Zambrini\dag\
and Stephen M. Barnett\dag }
\address{\dag\ Department of Physics, University of Strathclyde, Glasgow G4 0NG,
  UK \\
\ddag\ Department of Electronic \& Electrical Engineering, University College London, 
London WC1E 7JE, UK} 
\ead{joerg@phys.strath.ac.uk}

\begin{abstract}
The uncertainty relation for angle and angular momentum has a lower bound which depends on the form of the 
state. Surprisingly, this lower bound can be very large. We derive the states which have the lowest possible
uncertainty product for a given uncertainty in the angle or in the angular momentum. We show that, if 
the given angle uncertainty is close to its maximum value, the lowest possible uncertainty product 
tends to infinity.

\end{abstract}

\submitto{\JPB}
\pacs{03.65.Ta 40.25-p}


\maketitle


\section{Introduction}

If the lower bound in an uncertainty relation is state dependent, states satisfying the
equality in the uncertainty relation need not give a minimum in the uncertainty product. This contrasts
to the Heisenberg uncertainty principle for position and momentum \cite{heisenberg+:zphys:1927},
where the lower bound is a constant. Here, states which satisfy the equality in the uncertainty 
relation, that is intelligent states \cite{arragone}, also minimize the uncertainty product. The 
uncertainty relation for angular momentum and angle \cite{barnettpegg:pra41:1990} however, has 
a state dependent lower bound requiring a distinction between intelligent states and minimum 
uncertainty product states \cite{pegg+:njp7:2005}. These two kinds of states are defined as solutions of two 
different eigenvalue equations. For linear momentum and position the solutions to the two differential 
equations are the same Gaussians. In the angular case the two eigenvalue equations have two distinct solutions. 
Additionally, the angle is defined on a finite interval, allowing solutions to the differential equation 
which are disregarded in the linear case on the grounds that they do not represent physical, normalisable 
states on the infinite range on which position and linear momentum are defined. These 
solutions are peaked at the edges of the $2 \pi$ radian interval for the angle and consequently the 
angle uncertainty of these states tends to be larger than for cases 
where the wavefunction is peaked in the middle of the interval and decays towards the boundaries. 
The intelligent and minimum uncertainty product states thus appear in two varieties with small and 
large angle uncertainties. The distinction is most apparent for the uncertainty product, which is bounded
from above by $\hbar/2$ for the states with small angle uncertainty, but has no upper bound in the 
large-uncertainty case.

We should stress that in this work we will consider only the uncertainties in the angular momentum and
in the associated angular coordinate \cite{barnettpegg:pra41:1990}. This form of the uncertainty
relation has been demonstrated to hold in a recent experiment \cite{frankearnold+:njp6:2004} and underlies
the security of a lately developed free-space communication system \cite{gibson+:oe12:2004}. A range
of other uncertainty relations have been derived in which measures of angular spread other than the 
angle uncertainty are used. These includes uncertainties based on trigonometric functions of the 
angle \cite{trigonometricur}, discrete versions of the uncertainty relation \cite{discreteur} and 
entropic uncertainty relations \cite{entropicur}.

The family of states related to the angular uncertainty relation has been investigated in a series
of previous articles. The form of the angular uncertainty relation has been experimentally verified
using intelligent states with small angle uncertainties \cite{frankearnold+:njp6:2004}. The distinction
between intelligent states and minimum uncertainty states with small uncertainties has been presented
in a second article \cite{pegg+:njp7:2005}, where the possibility has been discussed to distinguish between
these in an experiment.
Intelligent states with large uncertainties have
been introduced in a third article \cite{goette+:job7:2005}, in which we have compared the analytically exact
expression for the wavefunction in terms of a special function with approximate expressions
for two limiting cases. The present article completes the study of this family of states by expounding
minimum uncertainty product states with large uncertainties. As the angle is defined on a 
finite interval its uncertainty is bounded from above and the uncertainty product reaches the global minimum for
eigenstates of the angular momentum operator $\hat{L}_z$ with uncertainty $\Delta L_z = 0$. But one can
also consider states which minimize the uncertainty product under the constraint of a given uncertainty in 
either the angle or the angular momentum. It has been shown that states minimizing the uncertainty product
are the same whether the additional constraint is a given uncertainty in the angle $\Delta \phi$ or a given
uncertainty in the angular momentum $\Delta L_z$ \cite{pegg+:njp7:2005}. These states are called constrained
minimum uncertainty product (CMUP) states. 

In this article we will extend the analysis of CMUP states to the large-uncertainty regime. An exact analytical
solution for the wavefunction of these states will be given in terms of complex confluent hypergeometric 
functions, but 
we also present an approximate solution in terms of Airy functions for the limiting case where the given angle
uncertainty $\Delta \phi$ tends to its upper bound $\pi$. In particular we will compare this kind of CMUP 
states with the large-uncertainty intelligent states.


\section{Angular uncertainty relation}
It is physically impossible to distinguish between two rotation angles differing by $2\pi$ radians. Within the
quantum mechanical description of rotation angles, this restricts angle eigenvalues to lie within a $2\pi$ radian
interval [$\theta_0, \theta_0 + 2\pi)$ \cite{barnettpegg:pra41:1990}. Choosing $\theta_0$ determines a particular
$2\pi$ radian interval and with it a particular angle operator $\hat{\phi}_\theta$ and hence a basis 
set of angle eigenstates. In the following we adhere to the choice of $\theta_0$ used in previous work 
\cite{pegg+:njp7:2005,frankearnold+:njp6:2004,goette+:job7:2005} by setting $\theta_0 = -\pi$ and dropping
the label on the angle operator $\hat{\phi}$. The lower bound in the general form of the uncertainty relation for 
two Hermitian operators is given by the expectation value of the commutator of these operators \cite{robertson+}. 
The commutator for angle and angular momentum operator is rigorously derived in a finite state space $\Psi$ of
$2L+1$ dimensions, spanned by the eigenstates $|m\rangle$ of the angular momentum operator $\hat{L}_z$ with
$m$ ranging from $-L, -L+1, \dots, L$ \cite{barnettpegg:pra41:1990}. Only after physical results have been 
calculated in the finite dimensional space $\Psi$, $L$ is allowed to tend to infinity. It is in this limit of
$L \to \infty$ that the expectation value of the commutator $[\hat{L}_z,\hat{\phi}]$ can be approximated
to an excellent degree for all physical preparable states, which results in the angular uncertainty relation
\begin{equation}
\label{eq:aur}
\Delta L_z \Delta \phi \geq \frac{1}{2} | 1 - 2\pi P(\pi) |.
\end{equation}
Here, we are using units in which $\hbar = 1$ and $P(\pi)$ is the angle probability density at the edge of the 
chosen $2\pi$ radian interval. This corresponds to our choice of $\theta_0 = -\pi$, as the probability density is 
periodic in the angle and $P(-\pi) = P(\pi)$. In general different states used to calculate the uncertainties 
$\Delta L_z$ and 
$\Delta \phi$ will have a different angle probability density $P(\pi)$ rendering the lower bound in the uncertainty
relation (\ref{eq:aur}) state dependent. From the uncertainty relation (\ref{eq:aur}) it is evident why angular 
momentum eigenstates give a global minimum for the uncertainty product. For an eigenstate of the angular momentum
operator the angle probability density takes on the constant value of $P(\phi) = 1/(2\pi)$ for $\phi$ in $[-\pi,\pi)$. 
The lower bound in the uncertainty relation thus is equal to zero and so is the uncertainty product as the angular
momentum uncertainty vanishes and the  angle uncertainty has the finite value of $\pi/\sqrt{3}$. This global minimum
in the uncertainty product is also the dividing point between small-uncertainty and large-uncertainty states. Away from 
this point the constrained minimum uncertainty product (CMUP) states give a minimum in the uncertainty product for a given 
$\Delta \phi$ or a given $\Delta L_z$.


\section{CMUP states}
\begin{figure}
\includegraphics{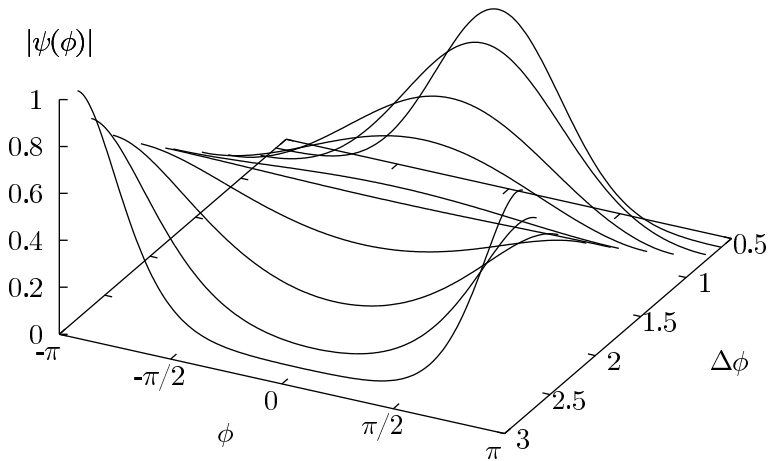}
\caption{\label{fig:wavefunction} Plot of the wavefunction of CMUP states showing the transition from the small-uncertainty
regime
to the large-uncertainty regime. This distinction refers to the angle uncertainty $\Delta \phi$, and the dividing point is
the flat wavefunction for $\Delta \phi = \pi / \sqrt{3}$.}
\end{figure}

Seeking states which minimize the uncertainty product for a given uncertainty in the angle or for a given uncertainty
in the angular momentum is equivalent to minimizing the uncertainty that is not given. The corresponding equation for 
CMUP states has been derived in \cite{pegg+:njp7:2005} using a variational method \cite{jackiw:jmp9:1968}. In this 
approach it is required that a CMUP state $| f \rangle$ minimizes the uncertainty product, but with the constraint of 
keeping the given variance constant and obeying the normalisation condition $\langle f | f \rangle = 1$. The 
additional constraints are taken into account by introducing undetermined Lagrange multipliers 
\cite{summypegg:oc77:1990}. In \cite{pegg+:njp7:2005} it has been shown that for a CMUP state $| f \rangle$ 
the mean values of angular momentum and angle can be set to zero, that is 
$\langle \hat{L}_z \rangle = \langle \hat{\phi} \rangle = 0$. Therefore, the variances $(\Delta L_z)^2$ and 
$(\Delta \phi)^2$ simplify to $\langle \hat{L}_z^2 \rangle$ and $\langle \hat{\phi}^2 \rangle$ respectively.
The variation of the uncertainty product with the given constraints thus results
in a linear combination of the variations $\delta \langle \hat{L}_z^2 \rangle, \delta \langle 
\hat{\phi}^2 \rangle$ and $\delta \langle f | f \rangle$, in which the Lagrange multipliers are the coefficients:
\begin{equation}
\label{eq:lincombvar}
\delta \langle \hat{L}_z^2 \rangle + \lambda \delta \langle \hat{\phi}^2 \rangle = \mu \delta \langle f | f \rangle.
\end{equation}
The linear combination is the same whether $\langle \hat{L}_z^2 \rangle$ is given and $\langle \hat{\phi}^2 \rangle$ 
is minimized or $\langle \hat{\phi}^2 \rangle$ is given and $\langle \hat{L}_z^2 \rangle$ is minimized.
Furthermore, it has also been demonstrated \cite{pegg+:njp7:2005} that it is admissible to consider only real 
coefficients $b_m$ in the 
angular momentum decomposition of $| f \rangle$ in the state space $\Psi$ 
\begin{equation}
| f \rangle = \sum_{m=-L}^{L} b_m | m \rangle.
\end{equation} 
This allows us to write the variation $\delta \langle f |\hat{A} | f \rangle$ for any Hermitian operator $\hat{A}$ as
$2 ( \delta \langle  f |) \hat{A} | f  \rangle$ \cite{summypegg:oc77:1990}. In particular this applies to the operators
$\hat{L}_z^2, \hat{\phi}^2$ and to the identity operator $\hat{I}$ corresponding to the variations
$\delta \langle \hat{L}_z^2 \rangle, \delta \langle \hat{\phi}^2 \rangle$ and $\delta \langle f | f \rangle$. The
linear combination of variations in (\ref{eq:lincombvar}) thus turns into a linear combination of operators applied 
to $| f \rangle$. This leads to an eigenvalue equation of the 
form
\begin{equation}
\label{eq:eigenvaleq}
\left( \hat{L}_z^2 + \lambda \hat{\phi}^2 \right) | f \rangle = \mu | f \rangle,
\end{equation}
where $\lambda$ and $\mu$ are the Lagrange multipliers. The identification of the angular momentum
operator $\hat{L}_z$ as derivative with respect to $\phi$ sets additional requirements on the
wavefunction representing CMUP states \cite{pegg+:njp7:2005,goette+:job7:2005}. The wavefunction in the 
angle representation $\psi(\phi) = \langle \phi | f \rangle$ has to be an element of $C^1$, which is the 
set of continuously differentiable functions. The question of differentiability 
is of particular 
importance at the boundaries of the $2\pi$ radian interval, on which the angle wavefunction is defined.
Whereas intelligent states are continuous, they do not have a continuous first derivative at $\phi = \pm \pi$.
CMUP states, however, do have a continuous first derivative at the boundaries, and therefore representing
$\hat{L}_z^2$ by the differential operator $-(\partial^2 / \partial \phi^2)$ is well defined. The eigenvalue
equation (\ref{eq:eigenvaleq}) may thus be turned into a differential equation for the CMUP wavefunction
$\psi(\phi)$:
\begin{equation}
\label{eq:cmupdiffeq}
\frac{\partial^2}{\partial \phi^2} \psi(\phi) = \left( \lambda \phi^2 - \mu \right) \psi(\phi).
\end{equation}
For the small-uncertainty case a solution of this equation has been given in terms of a 
confluent hypergeometric function
in reference \cite{pegg+:njp7:2005}. The angle wavefunction in this regime is peaked at $\phi = 0$ and 
$\lambda, \mu > 0$ such that the curvature of the wavefunction around $\phi = 0$ is negative. For the
large-uncertainty case the curvature around the central region is positive and the multipliers 
$\lambda, \mu < 0$. Formally we can give the solution to (\ref{eq:cmupdiffeq}) in terms of confluent 
hypergeometric functions with complex arguments:
\begin{equation}
\label{eq:comphypergeo}
\psi(\phi) \propto \exp\left( - \frac{\rmi |\lambda|^{\frac{1}{2}}}{2} \phi^2 \right) M \left(-\frac{\rmi}{4} \frac{|\mu|}
{|\lambda|^{\frac{1}{2}}} + \frac{1}{4},\frac{1}{2}, \rmi |\lambda|^{\frac{1}{2}}\phi^2 \right),  
\end {equation}
where $M$ is Kumer's function \cite{abrastegun:dov:1974}. This solution is obtained from (\ref{eq:cmupdiffeq}) by
setting $\lambda = -|\lambda|$ and $\mu = -|\mu|$ for $\lambda, \mu  < 0$ and using the same scaling as in 
\cite{pegg+:njp7:2005}, independent of the sign of $\lambda$ and $\mu$: $x = \sqrt{2} |\lambda|^{\frac{1}{4}} \phi$ and 
$a = |\mu|/(2|\lambda|^{\frac{1}{2}})$. With these substitutions (\ref{eq:cmupdiffeq}) takes on the form
\begin{equation}
\label{eq:scaleddiffeq}
\frac{\partial^2 \psi}{\partial x^2} + \left( \frac{x^2}{4} - a \right) \psi = 0,
\end{equation}
of which (\ref{eq:comphypergeo}) is a solution with the appropriate change of variables. To evaluate the wavefunction 
and to calculate the angle and angular momentum uncertainty we have solved the scaled differential equation 
(\ref{eq:scaleddiffeq}) numerically using a series expansion. In the scaled form the wavefunction is characterized
by the parameter $a$ which takes on positive values for large-uncertainty CMUP states. The appropriate scaling
is determined by the condition that the position $x_0$ of the first maximum of $\psi(x)$ corresponds to $\phi = \pi$,
such that
\begin{equation}
\label{eq:scaling}
\phi = \frac{x}{x_0} \pi \quad \mathrm{and} \quad  |\lambda| = \frac{x_0^4}{4\pi^4}.
\end{equation}
The value of $x_0$ is determined numerically, so that the scaled wavefunction can be normalised
between $-x_0$ and $x_0$ allowing a calculation of the expectation value $\langle \hat{x}^2 \rangle$ by numerical integration. 
The transition of the wavefunction for the CMUP states from the small-uncertainty regime to the large-uncertainty regime
is shown in figure \ref{fig:wavefunction}. 
In connection with figure \ref{fig:wavefunction} it is useful to discuss qualitatively the consequences of the positive 
curvature in the central region of the wavefunction for the large-uncertainty CMUP states on the angle uncertainty. 
In \cite{pegg+:njp7:2005} it
has been shown that for CMUP states with small $\Delta \phi$ the single-peaked wavefunction has approximately the form of 
a Gaussian. On increasing the scaling parameter $a$ towards zero the wavefunction becomes flater and deviates from the 
Gaussian form. For $a = 0$ the wavefunction is uniformly distributed between $-\pi$ and $\pi$ and the angle uncertainty 
takes on the value of $\Delta \phi = \pi / \sqrt{3}$. This is the dividing point between the small-uncertainty and 
large-uncertainty regime. If the parameter $a$ is further increased then the curvature turns positive and the wavefunction is peaked 
at $\pm \pi$. Calculating the uncertainty or for such a CMUP state from the variance $\Delta \phi^2 = \langle
\hat{\phi}^2 \rangle - \langle \hat{\phi} \rangle^2$ yields values for $\Delta \phi > \pi / \sqrt{3}$. In the limit of 
$a \to \infty$ the angle uncertainty approaches the maximum value $\pi$ \cite{goette+:job7:2005}. In this limit, the $2\pi$-periodic
angle probability distribution has a narrow peak centred at $\phi = \pi$. The width of this peak is \textit{not} $\Delta \phi$ but
rather the uncertainty in a different angle variable having a different range of allowed angles (for example $0$ to $2\pi$). The
angle uncertainty for a given state depends on this choice of allowed angles in precisely the same way as does the phase 
uncertainty for a harmonic osscillator or a single electromagnetic field mode \cite{phaseop}. 

\begin{figure}
\includegraphics{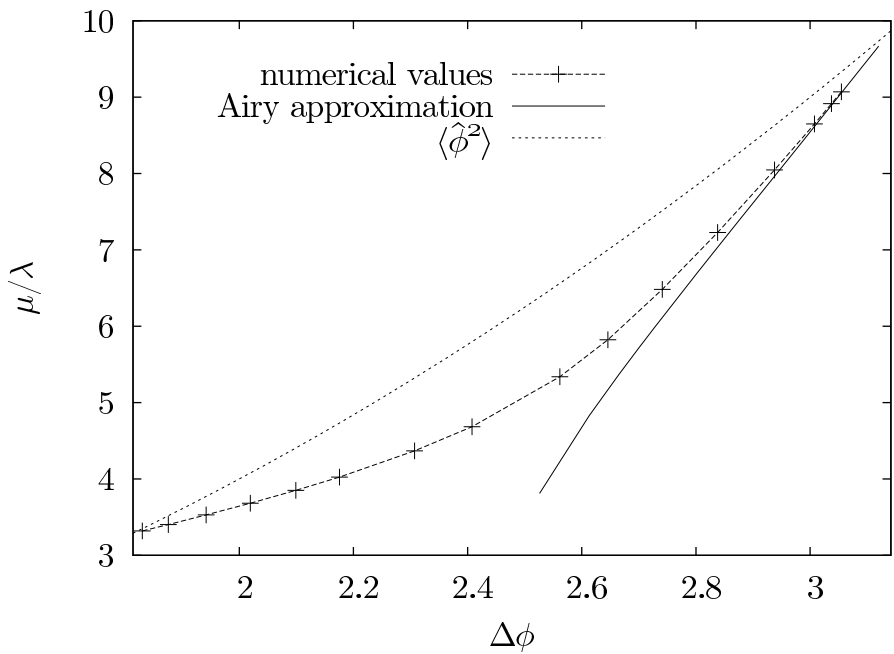}
\caption{\label{fig:mulambda} The ratio of the two Lagrange multipliers $\mu$ and $\lambda$
determines the limiting behaviour of the uncertainty product (see equation (\ref{eq:varianceprod})). For 
$\Delta \phi \to \pi$ the ratio $\mu/\lambda$ tends to $\pi^2$, but more slowly than $\langle \hat{\phi}^2
\rangle$. The uncertainty product thus tends to infinity. The plot of $\mu/\lambda$ in the Airy approximation
shows the region of validity for this approximation.}
\end{figure}

\begin{figure}
\includegraphics{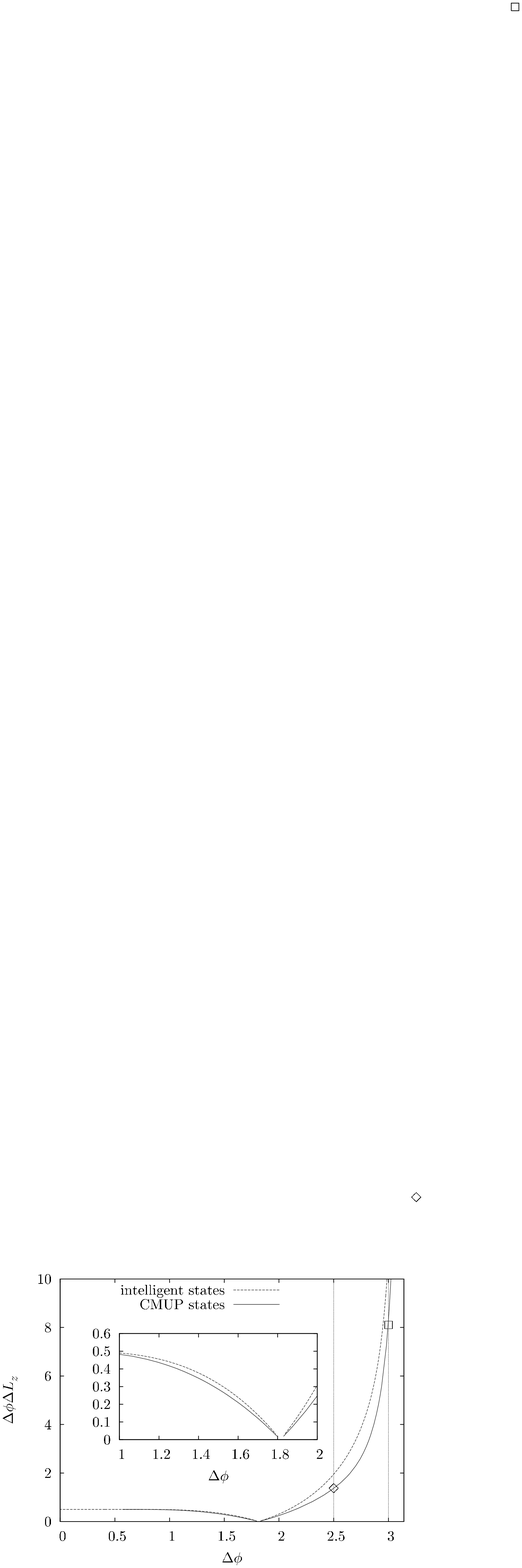}
\caption{\label{fig:uncertaintyprod} Plot of the uncertainty product as a function of $\Delta \phi$. The graphs
of the intelligent states \cite{frankearnold+:njp6:2004,goette+:job7:2005} and small-uncertainty CMUP states 
\cite{pegg+:njp7:2005} are shown for comparison (cf. plot of intelligent states in \cite{galindopascual:sv:1990}). 
The difference in the uncertainty product between intelligent states and CMUP states is significantly enhanced 
in the large-uncertainty regime for $\Delta \phi > \pi/\sqrt{3}$. For two values of $\Delta \phi$ (marked by the
dotted lines with the symbols $\Diamond$ and $\Box$) the difference in the wavefunction is shown in figure 
\ref{fig:intelcmupcomp}. The inset shows an enlargement around the global
minimum in the uncertainty product for $\Delta \phi = \pi/\sqrt{3}$. }
\end{figure}

\begin{figure}
\includegraphics{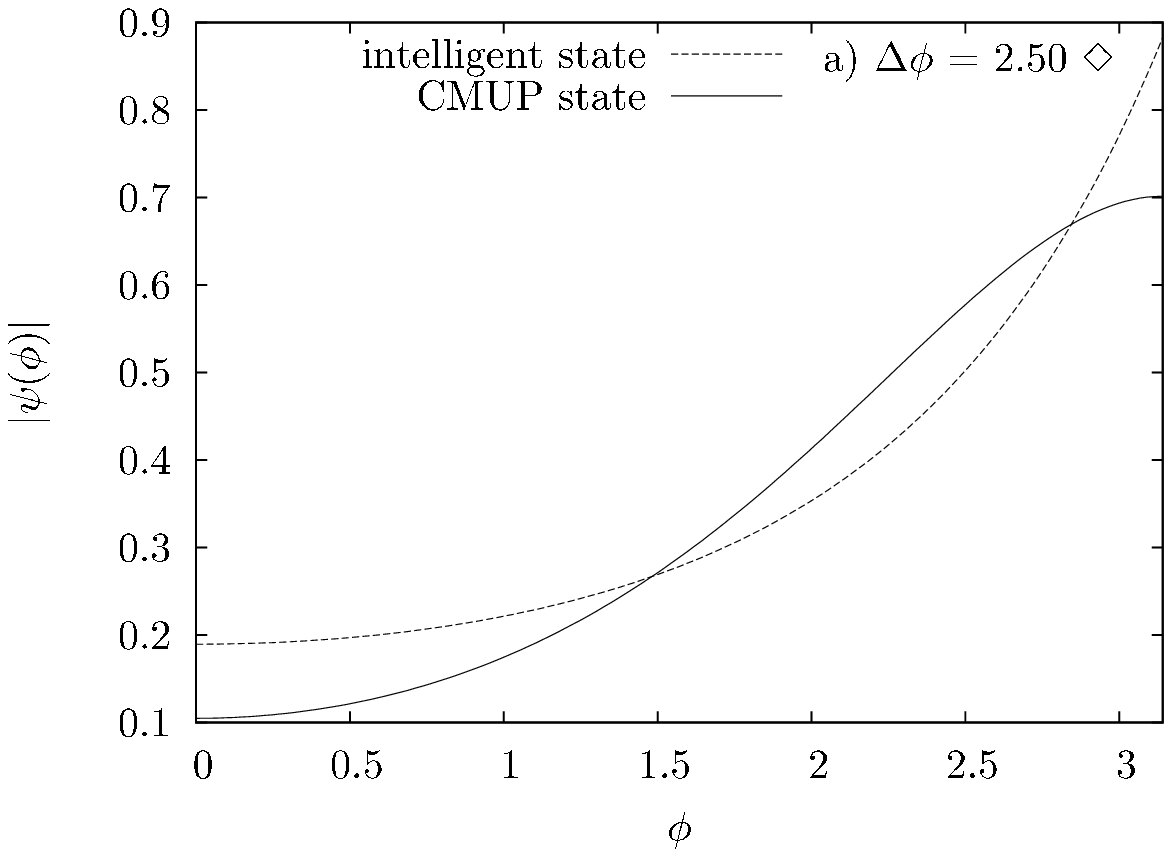} \\
\includegraphics{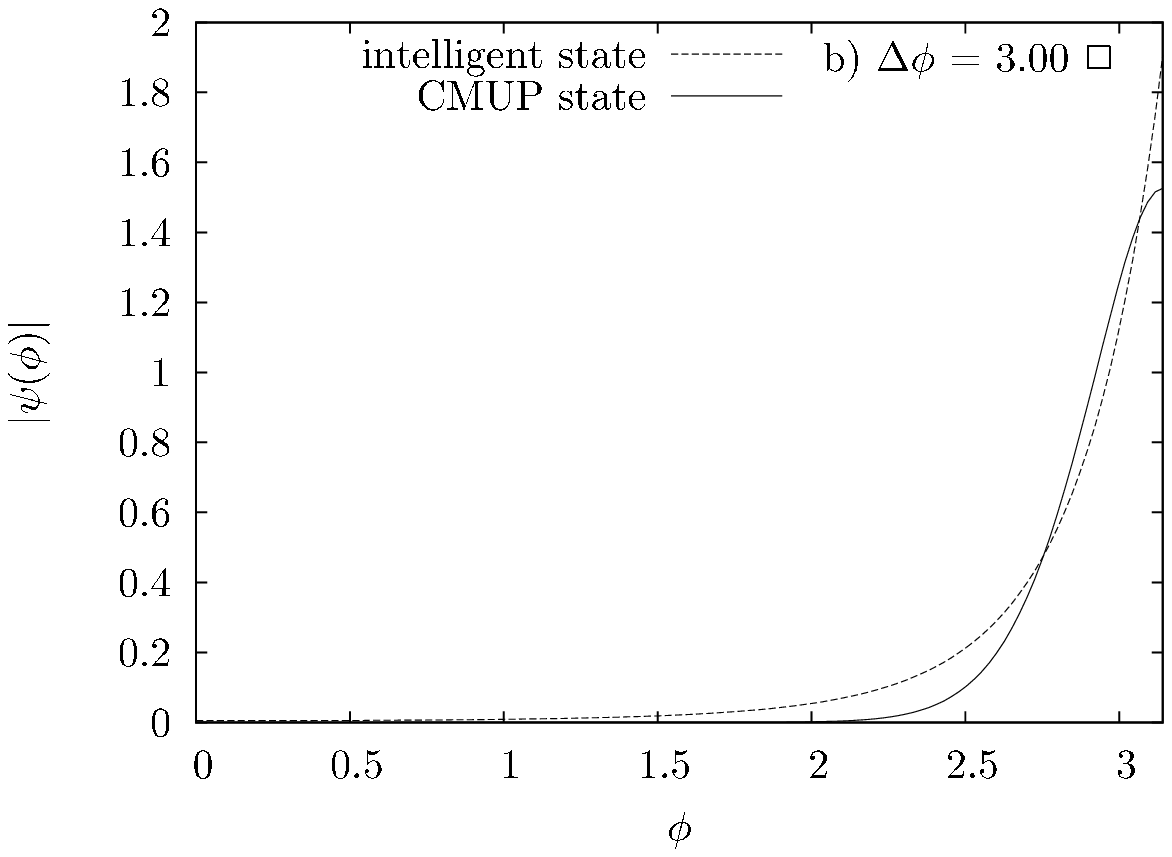}
\caption{\label{fig:intelcmupcomp} Comparison of the wavefunction for intelligent states and CMUP states for the same
$\Delta \phi$ in the large-uncertainty regime. The difference in the uncertainty product for the two values
$a)\: \Delta \phi = 2.5\: \Diamond$ and $b)\: \Delta \phi = 3.0\: \Box$ can be seen in figure \ref{fig:uncertaintyprod}. 
The position of the two values for $a)$ and $b)$ is marked in figure \ref{fig:uncertaintyprod}
by dotted lines with the respective symbols.}
\end{figure}

Using (\ref{eq:scaling}) the angle variance $\langle \hat{\phi}^2 \rangle$ is given by
$\langle \hat{\phi}^2 \rangle = (\pi^2 / x_0^2) \langle \hat{x}^2 \rangle$ for values of $x$ in $[-x_0,x_0)$.
The variance of the angular momentum operator is given by (\ref{eq:cmupdiffeq}) in terms of $\mu,\lambda$ and
$\langle \hat{\phi}^2 \rangle$, which results in the following expression for the product of the variances 
$\langle \hat{\phi}^2 \rangle$ and $\langle \hat{L}_z^2 \rangle$:
\begin{eqnarray}
\langle \hat{\phi}^2 \rangle \langle \hat{L}_z^2 \rangle & = \langle \hat{\phi}^2 \rangle \left( \mu - \lambda
\langle \hat{\phi}^2 \rangle \right),  \label{eq:varianceprod} \\
& = \langle \hat{x}^2 \rangle \left( -a + \frac{1}{4} \langle \hat{x}^2 \rangle \right).
\end{eqnarray}
The limiting behaviour of the uncertainty product is directly connected to 
the behaviour of the ratio
$\mu/\lambda$. For $a$ tending to zero, $\mu$ and $\lambda$ tend to zero individually but their ratio
$\mu/\lambda \to \pi^2/3$ (see figure \ref{fig:mulambda}). The variance $\langle \hat{\phi}^2 \rangle$ 
takes on the value of $\pi^2/3$ and the overall product of variances vanishes. For $a \to \infty$, 
both $\mu$ and  $\lambda \to -\infty$, but the ratio $\mu/\lambda$ approaches $\pi^2$. The variance 
$\langle \hat{\phi}^2 \rangle$,
however, tends to its maximum $\pi^2$ faster than $\mu/\lambda$ such that (\ref{eq:varianceprod}) tends to infinity.
The resulting behaviour of the uncertainty product as a function of $\Delta \phi$ is given in figure 
\ref{fig:uncertaintyprod}. As in the small-uncertainty case for $\Delta \phi < \pi/\sqrt{3}$ the 
uncertainty product is smaller for the CMUP states than for the intelligent states while still obeying 
the uncertainty relation (\ref{eq:aur}). This is possible because of the smaller probability density $P(\pi)$
at the edge of the chosen $2\pi$ radian interval. Also, the difference in the uncertainty product between
intelligent states and CMUP states in the large-uncertainty regime is enhanced over the small-uncertainty
regime. This goes along with a significant difference in the wavefunction for intelligent states and
CMUP states for the same $\Delta \phi$ in this region (see figure \ref{fig:intelcmupcomp}). 
In the small-uncertainty regime the wavefunction of intelligent and CMUP states both have approximately the 
same Gaussian form in the region where the uncertainty product is $1/2$ and changes only slowly with $\Delta \phi$.
In the large-uncertainty regime intelligent and CMUP states are of different form and we will discuss an approximate
expression for CMUP states in the limit of $\Delta \phi \to \pi$ later in this article. 

\begin{figure}
\includegraphics{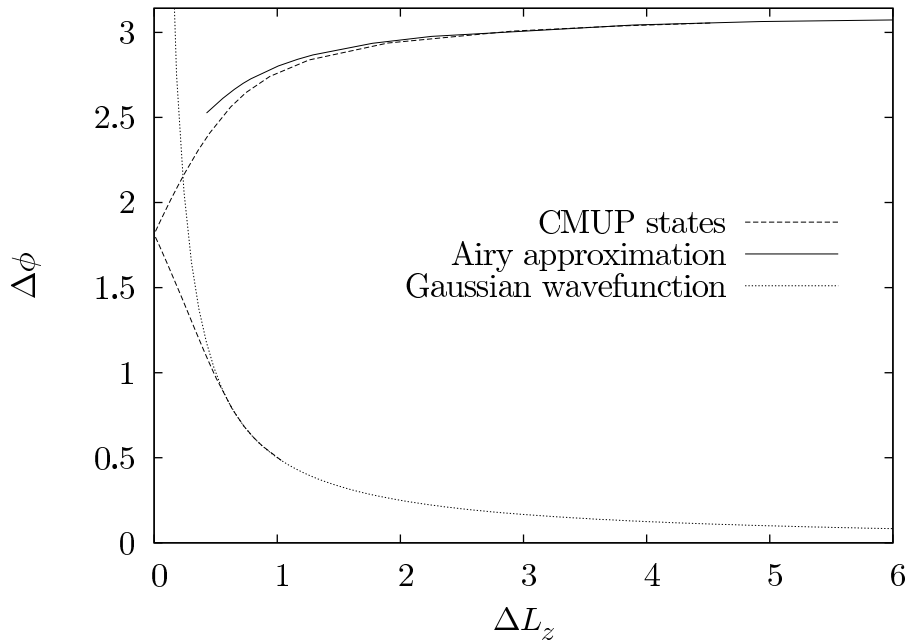}
\caption{\label{fig:deltaldeltaphi} If the uncertainty product is minimized for a given uncertainty in the angular momentum,
two minima can be obtained. The first and
smaller minimum is obtained for the small-uncertainty CMUP states while a secondary minimum is found for the large-uncertainty
CMUP states. For comparison the limiting cases for these two kinds of states are shown. For $\Delta \phi \to 0$ the small uncertainty
states become Gaussians \cite{pegg+:njp7:2005}, whereas the large-uncertainty states are approximatively given by Airy functions 
for $\Delta \phi \to \pi$.} 
\end{figure}

In connection with figure \ref{fig:uncertaintyprod} it is appropriate to clarify the meaning of minimizing the
uncertainty product under a constraint. For CMUP states with small and large angle uncertainties the angular momentum
uncertainty can take on all positive real values. $\Delta L_z$ is zero for the angular momentum eigenstates at 
$\Delta \phi = \pi/\sqrt{3}$ and it approaches infinity for $\Delta \phi \to 0$ and $\Delta \phi \to \pi$. Minimizing the
uncertainty product for a given $\Delta L_z$ yields two constrained minima. The smaller constrained minimum is obtained
for the small-uncertainty CMUP states and corresponds to an angle uncertainty $\Delta \phi < \pi/\sqrt{3}$. A secondary
minimum, however, is obtained for the large-uncertainty CMUP states corresponding to $\Delta \phi > \pi/\sqrt{3}$
(see figure \ref{fig:deltaldeltaphi}).
On the other hand minimizing the uncertainty product for a given $\Delta \phi$ results in a unique minimum. Whether this
minimum is obtained for small-uncertainty or large-uncertainty CMUP states depends on the given $\Delta \phi$.

Owing to the complexity of the CMUP states we are not able to give an analytical explanation of the
limiting behaviour in simple terms. Also, our method to determine the first maximum of the wavefunction 
numerically fails for very sharply peaked wavefunctions.  In the following we therefore present an approximate 
expression for the wavefunction  in terms of Airy functions, which allows us to calculate the variance
$\Delta \phi$ analytically.


\section{Airy approximation}
The defining differential equation for the CMUP states (\ref{eq:cmupdiffeq}) can be approximated and the
resulting equation solved to give an analytical expression for the CMUP wavefunction in the limiting
case $\Delta \phi \to \pi$, as we now describe. 

\begin{figure}
\includegraphics{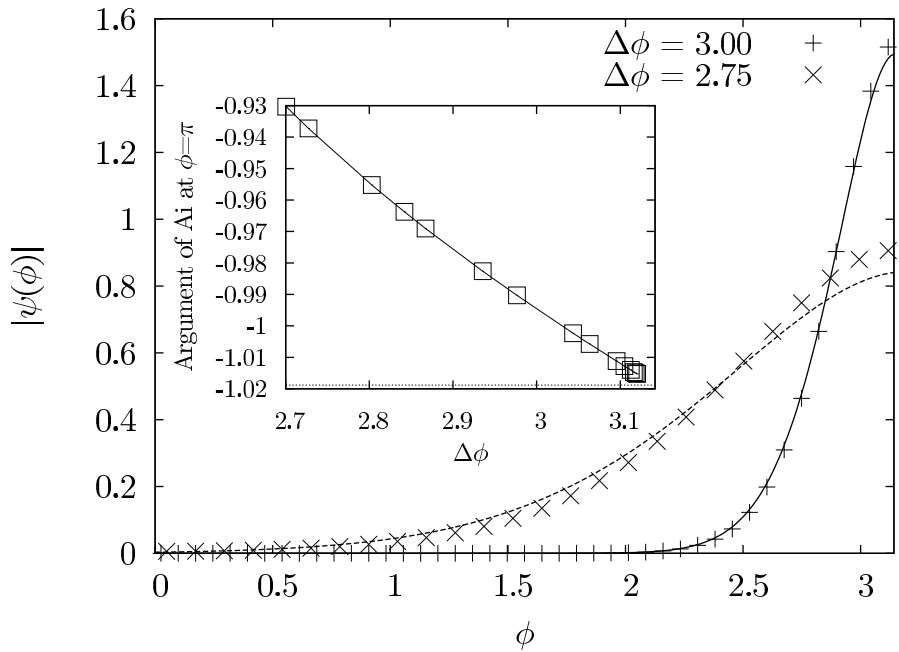}
\caption{\label{fig:wavefunccomp} Plot showing a comparison of the Airy approximation (continuous lines) with the 
numerical calculated wavefunction (individual points). For $\Delta \phi = 3$ ($+$) the Airy approximation shows
a good agreement with the numerical results. The inset shows the deviation of the argument of the Airy function 
$\Ai$ at $\phi = \pi$ from $-1.0188$ (marked by the horizontal dotted line), the position of the first maximum of the
Airy function $\Ai$.}
\end{figure}

The behaviour of the solution for a general differential equation of the form 
\begin{equation}
\label{eq:diffeq}
\frac{\partial^2 \psi}{\partial x^2} = P(x) \psi(x)
\end{equation}  
is partly determined by the sign of the function $P(x)$. Should $P(x)$ be purely positive we would expect
an exponential behaviour, whereas for a purely negative $P(x)$ the solution would be oscillating. Of particular
importance, therefore, are the values of $x$ where $P(x)$ exhibits a change of sign, that is the turning points of
the equation (\ref{eq:diffeq}). We can restrict the analysis of the differential equation (\ref{eq:cmupdiffeq}) to the half interval 
$[0,\pi)$ due to the symmetry of the equation. In this range equation (\ref{eq:cmupdiffeq}) has one turning
point at $\phi = \sqrt{\mu/\lambda}$. 
The equation is approximated by expanding $P(x) = \lambda \phi^2 - \mu$ around this turning point. Setting
$\phi = \sqrt{\mu/\lambda} + x$ and neglecting quadratic terms in $x$ turns (\ref{eq:cmupdiffeq}) into
Airy's equation \cite{airymethod}
\begin{equation}
\label{eq:airyeq}
\frac{\partial^2 \psi}{\partial y^2} = y \psi, \qquad y = -(2\sqrt{\mu \lambda})^{\frac{1}{3}}x = 
-(2\sqrt{\mu \lambda})^{\frac{1}{3}}(\phi - \sqrt{\mu/\lambda}).
\end{equation}
This equation is solved exactly by the Airy function $\Ai(y)$ which results in
\begin{equation}
\label{eq:airyfunc}
\psi(\phi) = C \Ai \left( - (2\sqrt{\mu \lambda})^{\frac{1}{3}} ( \phi - \sqrt{\mu/\lambda} ) \right)
\end{equation}
on substituting the appropriate variables.
Here, $C$ is the normalisation constant. To fulfill the boundary condition $\psi'(\pi) = 0$ the argument of 
the Airy function in (\ref{eq:airyfunc}) is required to have the value of the first zero of $\Ai'$ for $\phi=\pi$. 
This leads to the equation 
\begin{equation}
\label{eq:quartic}
-(2\sqrt{\mu \lambda})^{\frac{1}{3}} ( \pi - \sqrt{\mu/\lambda} ) = -1.0188.
\end{equation}
Choosing a particular $\lambda$ gives a quartic equation for $\sqrt{\mu/\lambda}$ and for values of 
$\sqrt{\mu/\lambda}$ close to $\pi$ an approximate solution is given by
\begin{equation}
\label{eq:quarticapprox}
\frac{\mu}{\lambda} \approx \pi - \left( \frac{-1.0188}{2\lambda\pi} \right)^{\frac{1}{3}}.
\end{equation}
In the Airy approximation a particular CMUP state can thus be characterized by the Lagrange multiplier
$\lambda$. The normalisation constant can be determined by analytical evaluation of the normalisation
integral
\begin{equation}
\label{eq:normint}
1 = 2 \int_0^\pi \psi^2(\phi) \rmd \phi \approx 2 C^2 \left( 2 \sqrt{\mu\lambda} \right)^{-\frac{1}{3}} \int_{-1.0188}^\infty 
\Ai^2(y) \rmd y.
\end{equation}
In the last step we have extended the range of integration from $y(\phi=0) = ( 2 \sqrt{\mu\lambda} )^{\frac{1}{3}}
\sqrt{\mu/\lambda}$ to infinity. In the region where the Airy approximation is applicable ($\Delta \phi \to \pi$), 
the wavefunction decays to zero sufficiently quickly for small angles so that extending the upper bound in 
the integral does not significantly change the normalisation integral.
Primitives of products of Airy functions can be calculated using the method of Albright \cite{albright}. This
results in 
\begin{equation}
C = \left( \mu \lambda \right)^{\frac{1}{12}} \left( (1.0188)^{\frac{1}{2}} (0.5357) 2^{\frac{1}{3}} \right)^{-1},
\end{equation}
where $\Ai(y=-1.0188) = 0.5357$. In figure \ref{fig:wavefunccomp} a comparison of the numerically calculated 
wavefunction and the wavefunction in the Airy approximation is shown. The approximation becomes better for 
values of $\Delta \phi$ closer  to $\pi$. An inset in figure \ref{fig:wavefunccomp} gives the deviation of the argument
of the Airy function in (\ref{eq:airyfunc}) from the exact value of $y = -1.0188$ due to the approximation
(\ref{eq:quarticapprox}) of the quartic equation (\ref{eq:quartic}) determining $\mu/\lambda$.

\begin{figure}
\includegraphics{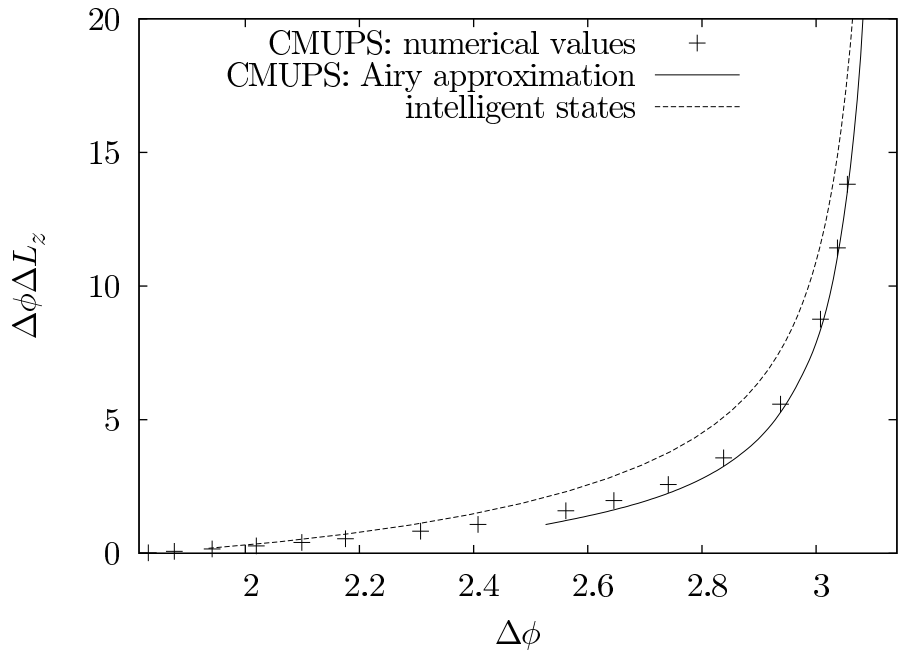}
\caption{\label{fig:numairycomp} Comparison of the uncertainty product calculated in the Airy approximation and for
numerical results. In difference to figure \ref{fig:uncertaintyprod}, the graph shows only the large-uncertainty 
region ($\Delta \phi > \pi/\sqrt{3}$) and the ordinate is extended to larger values of $\Delta \phi \Delta L_z$. 
The Airy approximation explains
the behaviour of $\Delta \phi \Delta L_z$ in a region where our numerical calculation fails.}
\end{figure}

Within the Airy approximation the integral for the angle variance can be calculated analytically using the 
method of Albright \cite{albright}:
\begin{eqnarray}
(\Delta \phi)^2 & = \frac{\mu}{\lambda} + \frac{2}{3} (1.0188) \left( 2 \sqrt{\mu \lambda} \right)^{-\frac{1}{3}} 
\sqrt{\mu/\lambda} \nonumber \\
& + \frac{1}{5} (1.0188^{-1} + 1.0188^2) \left( 2 \sqrt{\mu \lambda} \right)^{-\frac{2}{3}}.
\label{eq:airydeltaphi}
\end{eqnarray} 
As in the calculation of the normalisation constant (\ref{eq:normint}) the upper boundary in the integration
has been extended to infinity. On multiplying (\ref{eq:airydeltaphi}) by $\lambda$ one can see in 
(\ref{eq:varianceprod}) that $\lambda \langle \hat{\phi}^2 \rangle$ will always be smaller than $\mu$ resulting
in an unbounded uncertainty product. Within the Airy approximation the uncertainty product can be calculated
for values of $\Delta \phi$ much closer to $\pi$ than in the numerical calculation. This is due to the fact that
our numerical determination of the first maximum fails for large values of $a$. In the Airy approximation a numerical
search for the first maximum is not necessary. The uncertainty product calculated in the Airy approximation is compared
with the numerical results in figure \ref{fig:numairycomp}.


\section{Conclusion}
In this paper we have completed the study of states related to the angular uncertainty relation. This particular
uncertainty relation differs from the Heisenberg uncertainty principle for linear position and momentum in two ways.
The lower bound in the uncertainty relation is state dependent which causes intelligent states, that is
states satisfying the equality in the uncertainty relation, to be distinct from constrained minimum uncertainty product
(CMUP) states. These states minimize the uncertainty product for a given variance in angle or for a given variance in 
angular momentum. Also, in contrast to the linear position, the angle is defined on a bounded interval. Therefore,
wavefunctions peaked at the edge of the interval are normalisable and can represent physical states. Intelligent
and CMUP states are defined by two different eigenvalue equations. For the angular uncertainty relation the solutions
to the eigenvalue equations are angle wavefunctions which are peaked in the middle or peaked at the edge. This gives 
rise to two varieties of states with small and large angle uncertainties respectively.
Intelligent states with large angle uncertainties may have arbitrarily large uncertainty products while still satisfying
the equality in the uncertainty relation. Similarly, CMUP states with large-angle uncertainties minimize the uncertainty product 
locally or globally for a given constraint. The obtained minimum uncertainty product may be very large. It depends on the given 
uncertainty whether the minimum is local and the small uncertainty CMUP states give a smaller uncertainty product, or if the 
large-uncertainty
CMUP state gives the smallest possible constrained uncertainty product.
 
Intelligent states with small and large angle uncertainties have been discussed previously in two papers
\cite{frankearnold+:njp6:2004,goette+:job7:2005}, while the CMUP states with small angle uncertainties have been 
studied in a third paper 
\cite{pegg+:njp7:2005}. Here, we have examined CMUP states with large angle uncertainties. We have found an analytically exact
solution for the CMUP eigenvalue equation in terms of confluent hypergeometric functions with complex arguments. 
We also have solved the equation numerically and have calculated the angle uncertainty from this solution by numerical 
integration. To explain the limiting behaviour for sharply peaked wavefunctions we have developed an analytical
approximation using an expansion about a turning point. The approximate solution is given as the decaying tail of the 
Airy function $\Ai$. Within this approximation we were able to calculate the uncertainty product analytically.

We have found that the difference in the uncertainty product between intelligent states and CMUP state is enhanced
in the large-uncertainty regime.
In \cite{pegg+:njp7:2005} the possibility to distinguish between intelligent states and CMUP states in an experiment was
discussed. While it might be more difficult to prepare large-uncertainty states in an experiment, the greater difference 
in the uncertainty product could simplify the experimental evaluation significantly. The difference between CMUP and
intelligent states in the large-uncertainty regime is also a clear indication for the necessary distinction between the
two regimes and shows that large-uncertainty states cannot be transformed into small-uncertainty states by shifting the
the $2\pi$ radian range, but should be treated separately.


\ack
We would like to thank David T. Pegg for helpful discussions and we acknowledge
financial support from the UK Engineering and Physical Sciences Research Council 
(EPSRC) under the grant GR S03898/01.




\section*{References}
\begin{thebibliography}{99}
\bibitem{heisenberg+:zphys:1927} Heisenberg W 1927 \ZP
  \textbf{43} 127
\bibitem{arragone} Aragone C, Guerri G, Salam\'o S and Tani J L 1974 \JPA \textbf{7}
  L149
  \nonum Aragone C, Chalbaud E and Salam\'o S 1976 \textit{J. Math. Phys.} \textbf{17} 1963
\bibitem{barnettpegg:pra41:1990} Barnett S M and Pegg D T 1990 \textit{Phys. Rev. A}
  \textbf{41} 3427
\bibitem{pegg+:njp7:2005} Pegg D T, Barnett S M, Zambrini R, Franke-Arnold S and Padgett M 2005 \NJP \textbf{7} 62
\bibitem{frankearnold+:njp6:2004} Franke-Arnold S, Barnett S M, Yao E, Leach J,
  Courtial J and Padgett M 2004 \NJP \textbf{6} 103
\bibitem{gibson+:oe12:2004} Gibson G, Courtial J, Padgett M, Vasnetsov M, Pas'ko M, Barnett S M and
  Franke-Arnold S 2004 \textit{Opt. Express} \textbf{12} 5448
\bibitem{trigonometricur} Holevo A 1982 \textit{Probabilistic and statistical aspects of quantum theory}
  (Amsterdam: North Holland) 
  \nonum Yamada K 1982 \textit{Phys. Rev. D} \textbf{25} 3256
  \nonum Goh S S and Goodman T 2003 \textit{Appl. Comput. Harmon. Anal.} 
  \textbf{16} 19
\bibitem{discreteur} Brody D C and Meister B K 1999 \JPA \textbf{32} 4921
\bibitem{entropicur} Deutsch D 1983 \textit{Phys. Rev. Lett} \textbf{50} 631
  \nonum S\'anchez-Ruiz J 1993 \textit{Phys. Lett. A} \textbf{181} 193 
\bibitem{goette+:job7:2005} G\"otte J B, Zambrini R, Franke-Arnold S and Barnett S M 2005 \JOB \textbf{7} S563
\bibitem{robertson+} Robertson H P 1929 \textit{Phys. Rev} \textbf{34}
  163
 \nonum Schr\"odinger E 1930 \textit{Sitz. Preuss. Akad.
  Wiss.} \textbf{XIX} 296
\bibitem{jackiw:jmp9:1968} Jackiw R 1968 \textit{J. Math. Phys} \textbf{9} 339
\bibitem{summypegg:oc77:1990} Summy G S and Pegg D T 1990 \textit{Optics Communications} \textbf{77} 75
\bibitem{abrastegun:dov:1974} Abramowitz M and Stegun I A 1974 \textit{Handbook of Mathematical Functions}
  (Mineola: Dover Publications)
\bibitem{phaseop} Barnett S M and Pegg D T 1989 \textit{J. Mod. Opt} \textbf{36} 7
  \nonum Barnett S M and Pegg D T 1997 \textit{J. Mod. Opt} \textbf{44} 225 
\bibitem{galindopascual:sv:1990} Galindo A and Pascual P 1990
  \textit{Quantum Mechanics} vol 1 (Berlin: Spinger Verlag)
\bibitem{airymethod} Jeffreys H 1942 \textit{Phil. Mag} \textbf{33} 451
  \nonum Radmore P 1980 \JPA \textbf{13} 173
\bibitem{albright} Albright J R 1977 \JPA \textbf{19} 485 
  \nonum Vall\'ee O and Soares M 2004 \textit{Airy functions
  and applications to physics} (London: Imperial College Press)  
\end{thebibliography}
\end{document}